\documentclass[10pt,aps,showpacs,floatfix,twocolumn,tightenlines]{revtex4}
\usepackage{epsfig}
\usepackage{psfig}
\usepackage{graphicx}
\usepackage{graphics}
\usepackage{xspace}
\usepackage{amssymb}
\usepackage{amsmath}
\usepackage{latexsym}
\usepackage{natbib}
\usepackage{mathrsfs}
\newcommand{\inieq}{\begin{eqnarray}}            
\newcommand{\fineq}{\end{eqnarray}}            
\newcommand{\diff}{{\rm\,d}}                    
\newcommand{\bint}{\mskip .5mu \int \mskip-18mu} 
\newcommand{\half}{\textstyle{1\over 2}}

\def\s{\mbox{\bf s}}

\def\r{\mbox{{\bf  r}}}

\def\p{\mbox{\boldmath $p$}}

\def\q{\mbox{\boldmath $q$}}

\def\k{\mbox{\boldmath $k$}}

\def\ss{\mbox{\boldmath $\sigma$}}

\def\al{\mbox{\boldmath $\alpha$}}

\def\mcg{\mbox{$\mathcal{G}$}}
\def\mcv{\mbox{$\mathcal{V}$}}

\begin{document}
\title{Inclusive electron scattering in a relativistic Green function approach} 

\author{Andrea Meucci} 
\author{Franco Capuzzi}
\author{Carlotta Giusti}
\author{Franco Davide Pacati }
\affiliation{Dipartimento di Fisica Nucleare e Teorica, 
Universit\`{a} di Pavia and \\
Istituto Nazionale di Fisica Nucleare, 
Sezione di Pavia, I-27100 Pavia, Italy}

\date{\today}

\begin{abstract}
A relativistic Green function approach to the inclusive quasielastic $(e,e')$ 
scattering is presented. The single particle Green function is expanded
in terms of the eigenfunctions of the nonhermitian optical potential. This 
allows one 
to treat final state interactions consistently in the inclusive and 
in the exclusive reactions. Numerical results for the  response functions and
the cross sections for different target nuclei and in a wide range 
of kinematics are presented and discussed in comparison with experimental data. 
\end{abstract}

\maketitle

\section{Introduction}

The inclusive electron scattering in the quasielastic region addresses to the
one-body mechanism as a natural interpretation. However, when the experimental
data of the separation of the longitudinal and transverse responses were
available, it became clear that the explanation of both responses necessitated 
a more complicated frame than the single particle model coupled to one-nucleon 
knockout.

A review till 1995 of the experimental data and their possible explanations is
given in Ref. \cite{book}. Thereafter, only few experimental papers were 
published \cite{batesca,csfrascati}. 
New experiments with high experimental resolution are planned at 
JLab \cite{jlabpro}  in order to extract the response functions.

From the theoretical side,  a wide literature was produced in order to explain
the main problems raised by the the separation, i.e., the lack of strength in 
the longitudinal response and the excess of strength in the transverse one.
The more recent papers are mainly concerned with the contribution to the
inclusive cross section of meson
exchange currents and isobar excitations \cite{Sluys,Cenni,Amaro}, with
the effect of correlations \cite{Fabrocini,Co}, and the use
of a relativistic frame in the calculations \cite{Amaro}.

At present, however, the experimental data are not yet completely understood. 
A possible solution
could be the combined effect of two-body currents and tensor correlations
\cite{Leidemann,Fabrocini,Sick}.

In this paper we want to discuss the effects of final state interactions in a 
relativistic frame. Final state interactions are an important ingredient of the 
inclusive electron scattering, since they are essential to explain the 
exclusive one-nucleon emission, which gives the main contribution to the 
inclusive reaction in the quasielastic region. The absorption  
in a given final state due, e.g., to the imaginary part of the optical 
potential, produces a loss of flux that is appropriate for the exclusive 
process, but inconsistent for the inclusive one, where all the 
allowed final channels contribute and the total flux must be conserved. 

This conservation is preserved in the Green function approach considered here, 
where the components of the nuclear response are written in terms of the single
particle optical model Green function. This result was originally derived by
arguments based on the multiple scattering theory \cite{hori} and successively
by means of the Feshbach projection operator formalism 
\cite{chinn,bouch,capuzzi,capma}. The spectral representation of the
single particle Green function, based on a biorthogonal expansion in terms 
of the eigenfunctions of the nonhermitian optical potential, allows one to 
perform explicit calculations and to treat final state interactions 
consistently in the inclusive and in the exclusive reactions. Important 
and peculiar effects are given, in the inclusive reactions, by the imaginary 
part of the optical potential, which is responsible for the  redistribution of 
the strength among different channels. 

In a previous paper of ours \cite{capuzzi} the approach was used in a
nonrelativistic frame to perform explicit calculations of the longitudinal and 
transverse inclusive response functions.   
The main goal of this paper is to extend the method to a relativistic frame and
produce similar results. 
Although some differences and complications are due to the Dirac matrix 
structure, the formalism follows the same steps and approximations
as those developed in
the nonrelativistic frame of Refs. \cite{capuzzi,capma}.
The numerical results obtained in the relativistic approach allow us to check 
the relevance of relativistic effects in the kinematics already considered  
in Ref. \cite{capuzzi} and can be applied to a wider range of kinematics
where the nonrelativistic calculations are not reliable. 

In Sec. \ref{sec.green} the hadron tensor of the inclusive process is expressed 
in term of 
the relativistic Green function, which is reduced in Sec. \ref{sec.single} to 
a single particle expression. The problem of antisymmetrization is discussed in 
Sec. \ref{sec.anti}. In Sec. \ref{sec.spectral} the Green function is 
calculated in terms of the spectral representation related to the optical 
potential. In Sec. \ref{sec.results} the 
results of the calculation are reported and compared with the experimental 
data. Summary and conclusions are drawn in Sec. \ref{conc}.

\section{The Green function approach}
\label{sec.green}

\subsection{Definitions and main properties}

In the one photon exchange approximation the inclusive cross section for the
quasielastic $(e,e^{\prime})$ scattering on a nucleus is given by \cite{book}
\inieq
\sigma_{\textrm{inc}} = K \left( 2\varepsilon_{\textrm{L}}R_{\textrm{L}}
 + R_{\textrm{T}}\right) \ , \label{eq.cs}
\fineq 
where $K$ is a kinematical factor and 
\inieq
\varepsilon_{\textrm{L}} = \frac{Q^2}{\q^2} \left( 1 + 2 \frac{\q^2}{Q^2}
\tan^2{(\vartheta_e/2)}\right)^{-1} \ \label{eq.polar}
\fineq
measures the polarization of the virtual photon. In Eq. (\ref{eq.polar})
$\vartheta_e$ is the scattering angle of the electron, $Q^2 = \q^2 - \omega^2$,
and $q^{\mu} = (\omega,\q)$ is the four momentum transfer. All nuclear
structure information is contained in the longitudinal and transverse response
functions, $R_{\textrm{L}}$ and $R_{\textrm{T}}$, defined by
\inieq
R_{\textrm{L}}(\omega,q) &=& W_{\textrm{tot}}^{00}(\omega,q) \ , \nonumber \\
R_{\textrm{T}}(\omega,q) &=& W_{\textrm{tot}}^{11}(\omega,q) + 
      W_{\textrm{tot}}^{22}(\omega,q) \ ,
\label{eq.response}
\fineq
in terms of the diagonal components of the hadron tensor
\inieq
W^{\mu\mu}_{\textrm{tot}}(\omega,q) &=& 
 \bint\sum_{\textrm {f}} \mid \langle 
\Psi_{\textrm {f}}\mid J^{\mu}(\q) \mid \Psi_0\rangle\mid^2 \nonumber \\
&\times& 
\delta (E_0 +\omega - E_{\textrm {f}}) \ . \label{eq.hadrontensor}
\fineq
Here $J^{\mu}$ is the nuclear charge-current operator which connects the initial
state $\mid\Psi_0\rangle$ of the nucleus, of energy $E_0$, with the final states
$\mid \Psi_{\textrm {f}} \rangle$, of energy $E_{\textrm {f}}$, both
eigenstates of the $(A+1)$-body Hamiltonian $H$. The sum runs over the
scattering states corresponding to all of the allowed asymptotic configurations
and includes possible discrete states. As made for $\mid \Psi_{\textrm {f}} 
\rangle$, in the following the degeneracy indexes will be omitted whenever
unnecessary. The ground state $\mid\Psi_0\rangle$ is assumed to be 
nondegenerate. In order to avoid complications of little interest in the 
present context, we neglect recoil effects and consider only point-like 
nucleons, without distinguishing between protons and neutrons. Unless stated
otherwise, the wave functions are properly antisymmetrized.

The hadron tensor of Eq. (\ref{eq.hadrontensor}) can equivalently be expressed
as
\inieq
W^{\mu\mu}_{\textrm{tot}}(\omega,q) = -\frac{1}{\pi} \textrm{Im} \langle \Psi_0
\mid J^{\mu\dagger}(\q) G(E_{\textrm {f}}) J^{\mu}(\q) \mid \Psi_0 \rangle \
,\label{eq.hadrten}
\fineq
where $E_{\textrm {f}} = E_0 +\omega$ and $G(E_{\textrm {f}})$ is the Green
function related to $H$, i.e.,
\inieq
G(E_{\textrm {f}}) = \frac{1}{E_{\textrm {f}} - H + i\eta} \ . \label{eq.green}
\fineq
Here and in all the equations involving $G$ the limit for $\eta \rightarrow
+0$ is understood. It must be performed after calculating the matrix elements 
between normalizable states.

In this paper the interest is focused on relativistic wave functions for initial
and final states. Therefore, the $(A+1)$-body Hamiltonian $H$ is the sum of one
nucleon free Dirac Hamiltonians and two nucleons interactions $V_{jj'}$, i.e.,
\inieq
H = \sum_{j=1}^{A+1} \left( \al_j\cdot\p_j + \beta_jM\right) + 
\frac{1}{2} \sum_{j,j'=1}^{A+1} V_{jj'} \ , \label{eq.H}
\fineq
where the $4\times4$ Dirac matrices, $\al_j$ and $\beta_j$, act on the bispinor
variables of the nucleon $j$. No particular assumption is made on the $4\times4$ matrix
structure of $V_{jj'}$.

In order to express the hadron tensor in terms of single particle quantities,
the same approximations as in the nonrelativistic case \cite{capuzzi}
are required. The first one
consists in retaining only the one-body part of the charge-current operator
$J^{\mu}$. Thus, we set
\inieq
J^{\mu}(\q) = \sum_{i=1}^{A+1} j_i^{\mu}(\q) \ , \label{eq.1bcurrent}
\fineq
where $j_i^{\mu}$ acts only on the variables of the nucleon $i$. By Eq.
(\ref{eq.1bcurrent}), one can express the hadron tensor as the sum of two terms,
i.e.,
\inieq
W^{\mu\mu}_{\textrm{tot}}(\omega,q) = W^{\mu\mu}(\omega,q) +
W^{\mu\mu}_{\textrm{coh}}(\omega,q) \ , \label{eq.sumhadten}
\fineq
where $W^{\mu\mu}(\omega,q)$ is the incoherent hadron tensor \cite{west},
which contains only the diagonal contributions
$j_i^{\mu\dagger}Gj_i^{\mu}$, whereas the coherent hadron tensor 
$W^{\mu\mu}_{\textrm{coh}}(\omega,q)$ gathers the residual terms 
of interference between different nucleons. As the
incoherent hadron tensor, also $W^{\mu\mu}_{\textrm{coh}}(\omega,q)$ can be
expressed in terms of single particle quantities (see Sect. 9 of Ref.
\cite{capma}), but for the 
transferred momenta considered in this paper we can take advantage of
the high-$q$ approximation \cite{orlandini} and retain only 
$W^{\mu\mu}(\omega,q)$.
This term can be further simplified using the symmetry of $G$ for the exchange
of nucleons and the antisymmetrization of $\mid\Psi_0\rangle$. Therefore
we write
\inieq
W^{\mu\mu}_{\textrm{tot}}(\omega,q)  &\simeq& W^{\mu\mu}(\omega,q) 
=-\frac{A+1}{\pi}\nonumber \\
&\times& \textrm{Im}  \langle \Psi_0
\mid j^{\mu\dagger}(\q) G(E_{\textrm {f}}) j^{\mu}(\q) \mid \Psi_0 \rangle \ ,
\label{eq.apphadrten}
\fineq  
where $j^{\mu}(\q)$ is the component of $J^{\mu}(\q)$ related to an arbitrarily
selected nucleon. Due to the well-known completeness property, i.e.,
\inieq
\frac{1}{2 \pi i}\int \diff E \left( G^{\dagger}(E) - G(E)\right) = 1 \ ,
\label{eq.complete}
\fineq
the incoherent hadron tensor fulfills the energy sum rule
\inieq
\!\!\!\!\! \int \diff \omega W^{\mu\mu}(\omega,q) = (A+1) \langle \Psi_0
\mid j^{\mu\dagger}(\q) j^{\mu}(\q) \mid \Psi_0 \rangle  \ .
\label{eq.energysum}
\fineq


\subsection{Projection operator formalism}

This formalism yields an expression of the incoherent hadron tensor of 
Eq. (\ref{eq.apphadrten}) in terms
of eigenfunctions and Green functions of the optical potentials related to the
various reaction channels. Apart from complications due to the Dirac matrix
structure, we follow the same steps and approximations as in the 
nonrelativistic treatment \cite{capma,hori,chinn,capuzzi}.

Let us decompose $H$ as
\inieq
H = \al \cdot \p + \beta M + U + H_R \ , \label{eq.hdec}
\fineq
where $\al \cdot \p + \beta M$ is the kinetic energy of an arbitrarily selected
nucleon, $U$ is the interaction between this nucleon and the other ones, and
$H_R$ is the residual Hamiltonian of $A$ interacting nucleons. Such a 
decomposition cannot be performed in the physical space of the totally 
antisymmetrized $(A+1)$ nucleon wave functions. Therefore, we must operate in 
the Hilbert space $\mathcal{H}$ of the
wave functions which are antisymmetrized only for exchanges of the nucleons
of $H_R$. This treatment is presented here only for sake of simplicity. In Sect.
\ref{sec.anti} we shall discuss its physical drawbacks and outline the necessary
changes.

Let $\mid n\rangle$ and $\mid \varepsilon \rangle$ denote the antisymmetrized
eigenvectors of $H_R$ related to the discrete and continuous eigenvalues
$\varepsilon_n$ and $\varepsilon$, respectively. We introduce the 
operators $P_n$, projecting
onto the $n$-channel subspace of $\mathcal{H}$, and $Q_n$, projecting onto the
orthogonal complementary subspace, i.e.,
\inieq
P_n &=& \sum_a \int \diff \r \mid \r a ; n \rangle \langle n;\r a\mid \ ,
\nonumber \\
Q_n &=& 1 - P_n  \ . \label{eq.pro}
\fineq
Here $\mid \r a;n\rangle$ is the unsymmetrized vector obtained from the tensor
product between the discrete eigenstate $\mid n \rangle$ of $H_R$, and the
orthonormalized eigenvectors $\mid \r a\rangle$ $(a=1, 2, 3, 4)$ of the position
and the spin of the selected nucleon. The eigenvectors 
$\mid \r a\rangle$ have been chosen only for sake of definiteness, as every 
complete orthonormalized set of
single nucleon vectors would define the same operators $P_n$. Apart from minor
differences due to the present relativistic context, $P_n$ and $Q_n$ are the
projection operators of the Feshbach unsymmetrized formalism \cite{fesh}.
Note, for later use, the relations
\inieq
\left[ P_n, \al \cdot \p + \beta M \right] &=& 0 \nonumber \\
 H_R P_n &=& \varepsilon_n P_n \ . \label{eq.comm}
\fineq
Moreover, we introduce the projection operator onto the continuous channel
subspace, i.e.,
\inieq
P_c = \int \diff \varepsilon \sum_a \int \diff \r \mid \r a; \varepsilon \rangle 
\langle \varepsilon;\r a\mid \ . \label{eq.proc}
\fineq
Due to the completeness of the set $\left\{ \mid \r a ; n \rangle , 
\mid \r a ; \varepsilon \rangle \right\}$, one has
\inieq
\sum_n P_n + P_c = 1 \ . \label{eq.sumcom}
\fineq
Then, we insert Eq. (\ref{eq.sumcom}) into Eq. (\ref{eq.apphadrten}) 
disregarding the contribution of $P_c$. This approximation, which simplifies 
the calculations, is correct for sufficiently high values of the transferred
momentum $q$. Thus, the hadron tensor of Eq. (\ref{eq.apphadrten}) can be
expressed as the sum 
\inieq
W^{\mu\mu}(\omega,q) = W^{\mu\mu}_{\textrm{d}}(\omega,q) +
W^{\mu\mu}_{\textrm{int}}(\omega,q) \ , \label{eq.sumhad}
\fineq
of a direct term
\inieq
W^{\mu\mu}_{\textrm{d}}(\omega,q) &=& \sum_n W^{\mu\mu}_{n}(\omega,q) \ ,
\nonumber \\
W^{\mu\mu}_{n}(\omega,q) &=& -\frac{A+1}{\pi}\textrm{Im} \langle \Psi_0
\mid j^{\mu\dagger}(\q) P_n G(E_{\textrm {f}}) P_n \nonumber \\
&\times &j^{\mu}(\q) \mid \Psi_0 
\rangle \ , \label{eq.directterm}
\fineq   
and of a term
\inieq
W^{\mu\mu}_{\textrm{int}}(\omega,q) &=& 
\sum_n \widehat W^{\mu\mu}_{n}(\omega,q) \ ,\nonumber \\
\widehat W^{\mu\mu}_{n}(\omega,q) &=& -\frac{A+1}{\pi}\textrm{Im} \langle 
\Psi_0 \mid j^{\mu\dagger}(\q) P_n G(E_{\textrm {f}}) Q_n \nonumber \\ 
&\times &j^{\mu}(\q) \mid \Psi_0 \rangle \ , \label{eq.intterm}
\fineq   
which gathers the contributions due to the interference between the intermediate
states $\mid \r a;n\rangle$ related to different channels. 

We note that the interference term does
not contribute to the energy sum rule. In fact Eq. (\ref{eq.complete}) yields
\inieq
\int \diff \omega \ \widehat W^{\mu\mu}_{n}(\omega,q) &=& (A+1) \langle \Psi_0
\mid j^{\mu\dagger}(\q) P_n Q_n \nonumber \\ &\times & j^{\mu}(\q) \mid \Psi_0 
\rangle = 0 \ . \label{eq.completezero}
\fineq
Thus, the full contribution to the sum rule of the incoherent hadron tensor is
given only by the direct term, i.e.,
\inieq
\int \!\!&\diff \omega&\!\! W^{\mu\mu}(\omega,q) = \int \diff \omega \
W^{\mu\mu}_{\textrm{d}}(\omega,q) \nonumber \\ &=& (A+1) \langle \Psi_0
\mid j^{\mu\dagger}(\q) \sum_n P_nj^{\mu}(\q) \mid \Psi_0 \rangle \ ,
\label{eq.suminc}
\fineq
which, as a pure consequence of the omission of the continuous channels 
described by $P_c$, is smaller than the value of Eq. (\ref{eq.energysum}).

\section{Single particle expression of the hadron tensor}
\label{sec.single}

\subsection{Single particle Green functions}

For the time being, we disregard the effects of interference between different
channels and consider only the direct contribution to the hadron tensor of Eq.
(\ref{eq.directterm}). The matrix elements of $P_nG(E)P_n$ in the
basis $\mid \r a; n\rangle$ define a single particle Green function $\mcg_n(E)$ 
having a $4\times4$ matrix structure, i.e.,
\inieq
\langle \r a \mid \mcg_n(E) \mid \r ' a'\rangle \equiv \langle n;\r a\mid
G(E+\varepsilon_n)\mid \r ' a';n\rangle \ . \label{eq.matrixel}
\fineq
Note that here the energy scale is in accordance with Ref.
\cite{capma} and differs from Ref. \cite{capuzzi}.

The self-energy of $\mcg_n(E)$ is determined following the same steps used by 
Feshbach to determine the optical potential from the
Schr\"odinger equation \cite{fesh}. One starts from the relation
\inieq
\left( E - \al\cdot\p - \beta M - H_R - U + \varepsilon_n + i\eta \right) \left(
P_n + Q_n \right) \nonumber \\ 
\times G\left( E+\varepsilon_n\right) P_n = P_n \ , \label{eq.fesh}
\fineq
projects both sides by $P_n$ and then by $Q_n$, uses 
Eqs. (\ref{eq.comm}), resolves $Q_nGP_n$ in terms of $P_nGP_n$, and 
finally obtains
\inieq
\left( E - \al\cdot\p - \beta M - V_n(E) + i\eta \right)  \nonumber \\ 
\times P_n G(E+\varepsilon_n)
P_n = P_n \ , \label{eq.fesh2}
\fineq
with
\inieq
\!\!\!\!\!\!\!\! V_n(E) &=& P_nUP_n \nonumber \\ 
\!\!\!\!\!\!\!\! &+& P_nUQ_n \frac{1}{E-Q_nHQ_n+\varepsilon_n+i\eta}Q_nUP_n \ .
\label{eq.potenziale}
\fineq
Using Eq. (\ref{eq.pro}) for $P_n$ and considering the matrix elements in the
basis $\mid \r a; n\rangle$ of both sides of Eq. (\ref{eq.fesh2}), one has
\inieq
\mcg_n(E) = \frac{1}{E-h_n(E)+i\eta} \ , \label{eq.greenn}
\fineq
where
\inieq
h_n(E) = \al\cdot\p + \beta M + \mcv_n(E) \ , \label{eq.hn}
\fineq
and $\mcv_n(E)$ has the $4\times4$ matrix structure defined by
\inieq
\langle \r a \mid \mcv_n(E) \mid \r ' a'\rangle \equiv \langle n;\r a\mid
V_n(E)\mid \r ' a';n\rangle \ . \label{eq.matrixelv}
\fineq
Thus, $h_n(E)$ is the self-energy of the Green function $\mcg_n(E)$ and 
$\mcv_n(E)$ is
the related mean field. Using the same arguments as in the nonrelativistic case,
one finds that $\mcv_n(E)$ is the unsymmetrized Feshbach optical
potential \cite{fesh}, related to the channel $n$, for the relativistic
Hamiltonian $H$.

Using the first Eq. (\ref{eq.pro}) and Eq. (\ref{eq.matrixel}), the direct 
hadron tensor $W^{\mu\mu}_{n}(\omega,q)$ of Eq. (\ref{eq.directterm}) becomes
\inieq
W^{\mu\mu}_{n}(\omega,q) &=& -\frac{1}{\pi} \lambda_n \textrm{Im} \langle 
\varphi_n\mid j^{\mu\dagger}(\q) \mcg_n(E_{\textrm {f}}-\varepsilon_n) 
\nonumber \\
&\times &j^{\mu}(\q) \mid \varphi_n 
\rangle \ , \label{eq.directht}
\fineq
where the initial state $\mid \varphi_n \rangle$, normalized to $1$, is
represented by the bispinor defined by the matrix elements
\inieq
\langle\r a\mid\varphi_n\rangle \equiv \sqrt{\frac{A+1}{\lambda_n}} 
\langle n; \r a \mid \Psi_0\rangle \ , \label{eq.initial}
\fineq
$\lambda_n$ is the related spectral strength \cite{bofficapuzzi}
\inieq
\lambda_n = \left( A+1\right) \sum_a \int \diff \r \mid\langle n; \r a\mid 
\Psi_0\rangle \mid^2 \ , \label{eq.spst}
\fineq
with 
\inieq
\sum_n \lambda_n \simeq A + 1 \ ,
\fineq
and the symbol $\langle f\mid g\rangle$ denotes the scalar product
\inieq
\langle f\mid g\rangle = \sum_a\int \diff \r f^{\star}(\r a)g(\r a) \ .
\label{eq.scalarp}
\fineq

In Eq. (\ref{eq.directht}) the hadron tensor is expressed in terms of 
single particles quantities.
As in the nonrelativistic case, $\mid \varphi_n \rangle$ are the eigenstates of 
the optical potential, i.e.,
\inieq
\left( \al\cdot\p + \beta M + \mcv_n(E_0-\varepsilon_n)\right)\mid \varphi_n 
\rangle \nonumber \\ 
= (E_0-\varepsilon_n)\mid \varphi_n \rangle \ . \label{eq.eig}
\fineq
If $|\Psi_E\rangle$ is the eigenstate of $H$ corresponding
asymptotically to a nucleon, of momentum $\k$, colliding with a target nucleus in
the bound state $\mid \varepsilon_n\rangle$, the single particle vectors 
$\mid \chi_n (E-\varepsilon_n)
\rangle$ representing  the elastic scattering wave functions 
$\langle n; \r a\mid \Psi_E\rangle$ are eigenstates of the same optical 
potential, i.e.,
\inieq
\left( \al\cdot\p + \beta M + \mcv_n(E-\varepsilon_n)\right) \mid
\chi_n(E-\varepsilon_n)\rangle \nonumber \\ = (E-\varepsilon_n)\mid 
\chi_n(E-\varepsilon_n)\rangle   \ . \label{eq.elsc}
\fineq
Since $E$ is the total energy $\sqrt{k^2+M^2} + \varepsilon_n$, the argument
of $\mcv_n$ is the kinetic energy
(including the rest mass) of the emitted nucleon. 

\subsection{Interference hadron tensor}

The problem of expressing the interference hadron
tensor $\widehat W_n^{\mu\mu}$ in a one-body form is treated in Ref.
\cite{capuzzi} in the nonrelativistic context. It is argued that the
contribution of $\widehat W_n^{\mu\mu}$ can be included into the direct hadron
tensor $W_n^{\mu\mu}$ by the simple replacement
\inieq
\mcg_n(E) \rightarrow \mcg_n^{\textrm{eff}}(E) &\equiv& \sqrt{1-\mcv'_n(E)}
\mcg_n(E) \nonumber \\ 
&\times &\sqrt{1-\mcv'_n(E)} \ , \label{eq.simple}
\fineq
where $\mcv'_n(E)$ is the energy derivative of the Feshbach optical potential.

In Ref. \cite{capuzzi2} the problem is considered anew from a rigorous point of
view. The interference hadron tensor is expressed exactly as a series involving
energy derivatives of $\mcv_n(E)$, of increasing order, plus a residual term 
which
cannot be reduced to a single particle form. The series is expected to fastly 
converge near the quasielastic peak and at intermediate energies. It is argued
that in this region of momenta and energies the residual term is negligible.
Thus, one recovers the result of Eq. (\ref{eq.simple}) and second order 
corrections which do not seem to give a sizable contribution. 

Neither the treatment nor the
conclusions change if one considers the relativistic Hamiltonian $H$. Thus, for
the hadron tensor of Eq. (\ref{eq.sumhad}) we use the approximated expression 
obtained
from Eq. (\ref{eq.directht}) with the replacement (\ref{eq.simple}):
\inieq
W^{\mu\mu}(\omega , q) &=& -\frac{1}{\pi}\sum_n\lambda_n\textrm{Im} \langle 
\varphi_n
\mid j^{\mu\dagger}(\q) \mcg_n^{\textrm{eff}} (E_{\textrm {f}}-\varepsilon_n)
\nonumber \\
&\times &j^{\mu}(\q) \mid \varphi_n \rangle \ . \label{eq.replace}
\fineq   
Since the interference term $\widehat W_n^{\mu\mu}$ of 
Eq. (\ref{eq.intterm}) has no influence on the energy sum rule of the total 
hadron tensor of Eq. (\ref{eq.sumhad}), a natural question arises whether the
approximation leading to Eq. (\ref{eq.replace}) may change the sum 
rule. Actually,
one can observe that in Eq. (\ref{eq.simple}) $\mcg_n(E)$ is modified by factors
which change neither its properties of analyticity in the energy complex plane 
nor its high energy
behavior. This fact is used in Ref. \cite{capuzzi} to
prove the relation
\inieq
& &-\frac{1}{\pi} \int\diff E \,\, \textrm{Im} \langle \r a 
\mid \mcg_n^{\textrm{eff}}(E)
\mid \r ' a'\rangle \nonumber \\
 &=& -\frac{1}{\pi} \int \diff E \,\, \textrm{Im} \langle \r a \mid \mcg_n(E)
\mid \r ' a'\rangle \nonumber \\ 
&=& \delta \left(\r - \r '\right) \delta_{aa'} \ . \label{eq.prove}
\fineq
Therefore, the energy sum rule obtained from Eq. (\ref{eq.replace}) is
exactly the same as in Eq. (\ref{eq.suminc}), i.e., the correct sum rule of the
incoherent hadron tensor, apart from the contribution
of the continuous channels.

\subsection{Excited states of the residual nucleus}
As neither microscopic nor empirical calculations are available for the optical
potential $\mcv_n$ associated with the excited states 
$\mid \varepsilon_n\rangle$, a common practice relates them to the ground 
state potential $\mcv_0$ by means
of an appropriate energy shift. Here, as in Ref. \cite{capuzzi}, we use the
kinetic energy prescription for the shifts (see Sect. 5 of Ref. \cite{capma}),
naturally suggested by the plane wave impulse approximation. Such a prescription
preserves the value of the kinetic energy (including the rest mass), directly
related to the value of the optical potential variable in the energy scale
adopted here. Therefore we set
\inieq
\mcv_n(E) \simeq \mcv_0(E) \ , \label{eq.po}
\fineq
which implies
\inieq
\mcg_n(E) \simeq \mcg_0(E) \ . \label{eq.pog}
\fineq
Using these approximations in Eq. (\ref{eq.replace}), we write
\inieq
W^{\mu\mu}(\omega , q) &=& -\frac{1}{\pi}\sum_n \lambda_n\textrm{Im} 
\langle \varphi_n
\mid j^{\mu\dagger}(\q) \mcg_0^{\textrm{eff}} (E_{\textrm {f}}-\varepsilon_n)
\nonumber \\
&\times &j^{\mu}(\q) \mid \varphi_n \rangle \ . \label{eq.weff}
\fineq   
These approximations do not change the energy sum rule of 
$W^{\mu\mu}(\omega ,q)$.

\section{Antisymmetrization}
\label{sec.anti}

For sake of simplicity the treatment of Sects. \ref{sec.green} and 
\ref{sec.single} is based on the unsymmetrized projection operator $P_n$ defined
in Eq. (\ref{eq.pro}), leading to the Green function $\mcg_n$ of 
Eq. (\ref{eq.matrixel}). In this Section we examine the drawbacks of this
formulation and the possible alternatives. On the mathematical ground, $\mcg_n$
deserves the name of Green function since it fulfills the sum rule 
(\ref{eq.prove}), which is a qualifying property. Moreover, and intimately 
related, $\mcg_n$ is an invertible operator on the whole Hilbert 
space L$^2(\mathbb{R}^3)$.
Hence, its self-energy is not affected by any undue restriction of domain and by
the related mathematical troubles.

Notwithstanding, the optical potential $\mcv_n$ related to $\mcg_n$ suffers 
from the
drawback of having spurious eigenfunctions. In fact $H$ has both antisymmetrized
and unsymmetrized eigenvectors $\mid \Psi_E\rangle$ and the latter ones generate
eigenfunctions $\langle n;\r a\mid \Psi_E\rangle$ of $\mcv_n(E-\varepsilon_n)$ 
which have no physical meaning. No tool exists to make a distinction inside 
this unphysical degeneracy. Besides, it is apparent that $\mcv_n$ cannot be 
compared with any empirical optical model potential.

The remedy is a treatment based on projection operators onto antisymmetrized
states, although their inclusion into the hadron tensor is more laborious. Two
approaches are available.

{ a)} The first one (see Subsect. 3.2 of Ref. \cite{capma} and Appendix B of 
Ref. \cite{capuzzi}) uses the Feshbach projection operator onto antisymmetrized
states, i.e.,
\inieq 
P_n^{\textrm{F}} &=& \sum_{a,a'}\int\diff\r\diff\r '
a^{\dagger}_{{\mathbf {r}} a}\mid
n\rangle \nonumber \\
&\times & \langle\r a\mid\left(1-K_n\right)^{-1}\mid\r 'a'\rangle\langle n\mid
a_{\mathbf {r} 'a'}^{\protect{\phantom{\dagger}}} \ , \label{eq.proant}
\fineq
where $a^{\dagger}_{\mathbf {r} a}$ creates a nucleon in the state 
$\mid\r a\rangle$ and $K_n$ is the one-body density matrix defined by
\inieq
\langle\r a\mid K_n\mid\r 'a'\rangle \equiv \langle n\mid
a_{\mathbf{r'} a'}^{\dagger}
a_{\mathbf {r} a}^{\protect{\phantom{\dagger}}}
\mid n\rangle \ . \label{eq.density}
\fineq
The results of the previous Section remain true with the replacement
\inieq
& &\langle \r a \mid \mcg_n(E)\mid \r ' a'\rangle \rightarrow 
\langle \r a \mid \mcg_n^{\textrm{F}}(E)\mid \r ' a'\rangle \nonumber \\
&\equiv& \sum_b\int\diff \s 
\langle n\mid a_{\mathbf{r} a}^{\protect{\phantom{\dagger}}}
\frac{1}{E-H+\varepsilon_n+i\eta} a^{\dagger}_{\mathbf{s} b}\mid n\rangle 
\nonumber \\
&\times& \langle\s b\mid\left(1-K_n\right)^{-1}\mid\r 'a'\rangle \ ,
\fineq
where $\mcg_n^{\textrm{F}}$ is the Green function related to the {\lq\lq
symmetrized\rq\rq}  Feshbach optical 
potential $\mcv_n^{\textrm{F}}$ \cite{fesh2}.
In this approach the spurious degeneracy disappears, since one operates in a
Hilbert space of antisymmetrized states, but at the price of new drawbacks
\cite{capma,capma2}, namely: (1) $\mcg_n^{\textrm{F}}$ is not fully invertible 
and so in some cases it gives rise to incorrect Dyson equations; (2) both 
$\mcg_n^{\textrm{F}}$  and $\mcv_n^{\textrm{F}}$ are not symmetric for 
exchange $\r a\leftrightarrow \r ' a'$ and therefore $\mcv_n^{\textrm{F}}$ is 
nonhermitian below the threshold of the inelastic processes; (3) the usual 
nonlocal models of potential are probably inadequate 
for $\mcv_n^{\textrm{F}}$, which shows a complicate nonlocal structure.

In short, the approach based on $\mcv_n^{\textrm{F}}$ disguises nontrivial
mathematical problems and it is not really useful, since this potential bears 
no close relation with the empirical optical model potential.

{ b)} The second approach, where the above drawbacks disappear, is the one of 
Ref. \cite{capma}. It is based on the extended projection operator 
of Ref. \cite{capma3}
\inieq
P_n^{(p+h)} &=& \sum_a\int\diff\r \left(a^{\dagger}_{\mathbf{r}a} +
a_{\mathbf{r} a}^{\protect{\phantom{\dagger}}}\right) 
\mid n\rangle \nonumber \\ 
&\times& \langle n\mid \left(a^{\dagger}_{\mathbf{r} a} + 
a_{\mathbf{r} a}^{\protect{\phantom{\dagger}}}
\right) \ ,
\label{eq.procap}
\fineq
which leads to 
\inieq
& &\langle \r a \mid \mcg_n(E)\mid \r ' a'\rangle \rightarrow 
\langle \r a \mid \mcg_n^{(p+h)}(E)\mid \r ' a'\rangle \nonumber \\
&\equiv& \langle n\mid a_{\mathbf{r} a}
\frac{1}{E-H+\varepsilon_n+i\eta} a^{\dagger}_{\mathbf{r'} a'}\mid n\rangle 
\nonumber \\
&+& \langle n\mid a^{\dagger}_{\mathbf{r'} a'}\frac{1}{E+H-\varepsilon_n-i\eta} 
a_{\mathbf{r} a}\mid n \rangle \ . \label{eq.replace2}
\fineq
$\mcg_n^{(p+h)}(E)$ is the full Green function, including particle and
hole contributions. It fulfills the sum rule of Eq. (\ref{eq.prove}), is fully
invertible and produces mathematically correct Dyson equations. The related mean
field $\mcv_n^{(p+h)}(E)$ has no spurious eigenfunctions corresponding to
unsymmetrized states and its properties of nonlocality and symmetry make it more
easily comparable with the empirical optical model potentials. 
Therefore, we understand that in the following equations $\mcg_n$ will denote 
the full Green function $\mcg^{(p+h)}_n$ of the relativistic Hamiltonian $H$.
The associated mean field $\mcv_n$ is nonlocal as in the nonrelativistic case 
and does not conserve the primarily 4$\times$4 matrix structure of $V_{jj'}$.


\section{Spectral representation of the hadron tensor}
\label{sec.spectral}

In this Section we consider the spectral representation of the single particle
Green function which allows practical calculations of the hadron tensor of Eq.
(\ref{eq.weff}). In expanded form, it reads
\inieq
W^{\mu\mu}(\omega , q) &=& -\frac{1}{\pi} \sum_n \lambda_n \textrm{Im} 
\langle \varphi_n
\mid j^{\mu\dagger}(\q) \sqrt{1-\mcv'(E)}
\nonumber \\
&\times & \mcg(E)\sqrt{1-\mcv'(E)}j^{\mu}(\q) \mid \varphi_n \rangle \ , 
\label{eq.exweff}
\fineq
where $E=E_{\textrm {f}}-\varepsilon_n$. Here and below, the lower index $0$ is
omitted in the Green functions and in the related quantities. According to the
discussion of the previous Section, we understand that $\mcg$ is the full
particle-hole Green function of Eq. (\ref{eq.replace2}) and that $\mcv$ is the 
mean field potential related to $\mcg$ by the equations
\inieq
\mcg(E) &=& \frac{1}{E-h(E)+i\eta} \ , \label{eq.fullG}\\
h(E) &=& \al\cdot\p + \beta M + \mcv(E) \ .
\fineq
The use of this Green function does not change the expressions of the normalized
initial states $\mid\varphi_n\rangle$ and of the related spectroscopic factors
$\lambda_n$, defined in Eqs. (\ref{eq.initial}) and (\ref{eq.spst}),
respectively. Equivalently, they can be written as
\inieq
\langle\r a\mid\varphi_n\rangle &=& \lambda_n^{-\half} \langle n\mid
a_{\mathbf{r} a}
\mid \Psi_0\rangle \ , \label{eq.famla} \\
\lambda_n &=& \sum_a \int \diff \r \mid \langle n\mid a_{\mathbf{r} a}
\mid \Psi_0 \rangle \mid ^2 \ . \label{eq.famla2}
\fineq
Due to the complex nature of $\mcv(E)$ the spectral representation of $\mcg(E)$
involves a biorthogonal expansion in terms of the eigenfunctions of $h(E)$ and
$h^{\dagger}(E)$. We consider the incoming wave scattering solutions of the
eigenvalue equations, i.e.,
\inieq 
\left(\mathcal{E} - h^{\dagger}(E)\right) \mid
\chi_{\mathcal{E}}^{(-)}(E)\rangle &=& 0 \ , \label{eq.inco1} \\ 
\left(\mathcal{E} - h(E)\right) \mid
\tilde {\chi}_{\mathcal{E}}^{(-)}(E)\rangle &=& 0 \ . \label{eq.inco2}
\fineq
The choice of incoming wave solutions is not strictly necessary, but it is
convenient in order to have a closer comparison with the treatment of the
exclusive reactions, where the final states fulfill this asymptotic
condition and are the eigenfunctions
$\mid\chi_E^{(-)}(E)\rangle$ of $h^{\dagger}(E)$.

The eigenfunctions of Eqs. (\ref{eq.inco1}) and (\ref{eq.inco2}) satisfy the
biorthogonality condition
\inieq
\langle\chi_{\mathcal{E}}^{(-)}(E)\mid
\tilde {\chi}_{\mathcal{E}'}^{(-)}(E)\rangle = \delta 
\left(\mathcal{E} - \mathcal{E}' \right) \ , \label{bicon}
\fineq
and, in absence of bound eigenstates, the completeness relation
\inieq
\int_M^{\infty} \diff \mathcal{E}\mid\tilde
{\chi}_{\mathcal{E}}^{(-)}(E)\rangle\langle\chi_{\mathcal{E}}^{(-)}(E)\mid =1 
\ , \label{eq.comple}
\fineq
where the nucleon mass $M$ is the threshold of the continuum of $h(E)$. 

Eqs. (\ref{bicon}) and (\ref{eq.comple}) are the mathematical basis for 
the biorthogonal expansions. The contribution of possible bound state solutions 
of Eqs. (\ref{eq.inco1}) and (\ref{eq.inco2}) can be disregarded in 
Eq. (\ref{eq.comple})  
since their effect on the hadron tensor is negligible at the energy and momentum
transfers considered in this paper.

Inserting Eq. (\ref{eq.comple}) into Eq. (\ref{eq.fullG}) and using Eq.
(\ref{eq.inco2}), one obtains the spectral representation
\inieq
\mcg(E) &=& \int_M^{\infty} \diff \mathcal{E}\mid\tilde
{\chi}_{\mathcal{E}}^{(-)}(E)\rangle \nonumber \\
&\times& \frac{1}{E-\mathcal{E}+i\eta} \langle\chi_{\mathcal{E}}^{(-)}(E)\mid 
\ . \label{eq.sperep}
\fineq
Therefore, Eq. (\ref{eq.exweff}) can be written as
\inieq
W^{\mu\mu}(\omega , q) &=& -\frac{1}{\pi} \sum_n  \textrm{Im} \bigg[
 \int_M^{\infty} \diff \mathcal{E} \frac{1}{E_{\textrm
{f}}-\varepsilon_n-\mathcal{E}+i\eta}  \nonumber \\
&\times&  T_n^{\mu\mu}(\mathcal{E} ,E_{\textrm{f}}-\varepsilon_n) \bigg]
\ , \label{eq.pracw}
\fineq
where
\inieq
T_n^{\mu\mu}(\mathcal{E} ,E) &=& \lambda_n\langle \varphi_n
\mid j^{\mu\dagger}(\q) \sqrt{1-\mcv'(E)}
\mid\tilde{\chi}_{\mathcal{E}}^{(-)}(E)\rangle \nonumber \\
&\times& \!\! \langle\chi_{\mathcal{E}}^{(-)}(E)\mid  \sqrt{1-\mcv'(E)} j^{\mu}
(\q)\mid \varphi_n \rangle  \ . \label{eq.tprac}
\fineq
The limit for $\eta \rightarrow +0$, understood before the integral of Eq.
(\ref{eq.pracw}), can be calculated exploiting the standard symbolic relation
\inieq
\lim_{\eta \rightarrow 0} \frac{1}{E-\mathcal{E}+i\eta} = \mathcal{P}
\left(\frac{1}{E-\mathcal{E}}\right) - i \pi \delta \left(E-\mathcal{E}\right) 
\ , \label{eq.princ}
\fineq
where $\mathcal{P}$ denotes the principal value of the integral. Therefore, Eq.
(\ref{eq.pracw}) reads
\inieq
W^{\mu\mu}(\omega , q) = \sum_n \Bigg[ \textrm{Re} T_n^{\mu\mu}
(E_{\textrm{f}}-\varepsilon_n, E_{\textrm{f}}-\varepsilon_n)  \nonumber
\\
- \frac{1}{\pi} \mathcal{P}  \int_M^{\infty} \diff \mathcal{E} 
\frac{1}{E_{\textrm{f}}-\varepsilon_n-\mathcal{E}} 
\textrm{Im} T_n^{\mu\mu}
(\mathcal{E},E_{\textrm{f}}-\varepsilon_n) \Bigg] \ , \label{eq.finale}
\fineq
which separately involves the real and imaginary parts of $T_n^{\mu\mu}$.

Some remarks on Eqs. (\ref{eq.tprac}) and (\ref{eq.finale}) are in order.  
Let us examine the expression of $T_n^{\mu\mu}(\mathcal{E},E)$ at 
$\mathcal{E}=E=E_{\textrm{f}}-\varepsilon_n$ for a fixed $n$. This is the most 
important case since it appears in the first term in the right hand side of 
Eq. (\ref{eq.finale}), which gives the main contribution. 
Disregarding the square root correction, due to interference effects, one
observes that in Eq. (\ref{eq.tprac}) the second matrix element (with the
inclusion of $\sqrt{\lambda_n}$) is the transition amplitude for the single
nucleon knockout from a nucleus in the state $\mid \Psi_0\rangle$ leaving the
residual nucleus in the state $\mid n \rangle$. The attenuation of its strength,
mathematically due to the imaginary part of $\mcv^{\dagger}$, is related to the
flux lost towards the channels different from $n$. In the inclusive response
this attenuation must be compensated by a corresponding gain due to the flux
lost, towards the channel $n$, by the other final states asymptotically
originated by the channels different from $n$. In the description provided by
the spectral representation of Eq. (\ref{eq.finale}), the compensation is
performed by the first matrix element in the right hand side of 
Eq. (\ref{eq.tprac}), where the imaginary part of $\mcv$ has the effect of 
increasing
the strength. Similar considerations can be made, on the purely mathematical
ground, for the integral of Eq. (\ref{eq.finale}), where the 
amplitudes involved in $T_n^{\mu\mu}$ have no evident physical meaning as 
$\mathcal{E}\neq E_{\textrm{f}}-\varepsilon_n$. 
We want to stress
here that in the Green function approach it is just the imaginary part of 
$\mcv$ which accounts for the redistribution of the strength among different 
channels.

The matrix elements in Eq. (\ref{eq.tprac}) contain the mean field $\mcv(E)$ 
and its hermitian conjugate $\mcv^{\dagger}(E)$, which are nonlocal operator 
with a possibly complicated matrix structure. Neither microscopic nor empirical 
calculations of $\mcv(E)$ are available. In contrast, phenomenological optical 
potentials are available. They are obtained from fits to experimental data, are 
local and involve scalar and vector components only. The necessary replacement 
of the mean field by the empirical optical model potential is, however, a 
delicate step.

In the nonrelativistic treatment of Refs. \cite{capuzzi,capuzzi2} this
replacement is justified on the basis of the approximated equation (holding for
every state $\mid \psi\rangle$)
\inieq
& &  \textrm{Im} \langle \psi\mid \sqrt{1-\mcv'(E)} \mcg(E) 
\sqrt{1-\mcv'(E)} \mid \psi\rangle \nonumber \\ 
 &\simeq& \! \! \! 
\textrm{Im} \langle \psi\mid \sqrt{1-\mcv_{\textrm {L}}'(E)} 
 \mcg_{\textrm {L}}(E) \sqrt{1-\mcv_{\textrm {L}}'(E)} \mid \psi\rangle \ ,\,
\label{eq.approximate}
\fineq
where $\mcv_{\textrm {L}}(E)$ is the local phase-equivalent potential identified 
with the phenomenological optical model potential and $\mcg_{\textrm {L}}(E)$
is the related Green function. 
In Ref. \cite{capuzzi2} the proof of  Eq. (\ref{eq.approximate}) is based on 
two reasons: (1) a model of $\mcv(E)$ commonly used in dispersion relation
analyses; (2) the combined effect of the factor $\sqrt{1-\mcv'(E)}$ and
of the Perey factor, which connects the eigenfunctions of $\mcv(E)$ and 
$\mcv_{\textrm {L}}(E)$. We stress that it is just the factor 
$\sqrt{1-\mcv'(E)}$, introduced to account for interference 
effects, which allows the replacement of $\mcv(E)$ by $\mcv_{\textrm {L}}(E)$. 

Although the Perey effect is not sufficiently known for the Dirac equation, we 
have a reasonable confidence that Eq. (\ref{eq.approximate}) holds also in the 
present relativistic context. Therefore, we insert Eq. (\ref{eq.approximate}) 
into Eq. (\ref{eq.exweff}). Then, all the developments of this Section 
can be repeated with the simple replacement of $\mcv(E)$ by 
$\mcv_{\textrm {L}}(E)$. 

\section{Results and discussion} \label{sec.results}

The cross sections and the response functions of the inclusive quasielastic
electron scattering are calculated from the single particle expression of the
coherent hadron tensor in  Eq. (\ref{eq.finale}). After the replacement of the 
mean field $\mcv(E)$ by the empirical optical model potential 
$\mcv_{\textrm {L}}(E)$, the matrix elements of the nuclear current operator in 
Eq. (\ref{eq.tprac}), which represent the main ingredients of the calculation, 
are of the same kind as those giving the transition amplitudes  of the electron 
induced nucleon knockout reaction in the relativistic distorted wave impulse 
approximation (RDWIA)~\cite{meucci1}. Thus, the same treatment can be used 
which was successfully applied to describe exclusive $(e,e^{\prime}p)$ and 
$(\gamma,p)$ data~\cite{meucci1,meucci2}.

The final wave function is written in terms of its upper
component following the direct Pauli reduction scheme, i.e.,
\inieq
\!\!\!\! \chi_{\cal{E}}^{(-)}(E) \!\! = \!\! \left(\begin{array}{c} 
{\displaystyle \Psi_{\textrm {f}+}} \\ 
\frac{\displaystyle 1} {\displaystyle 
M+{\cal E}+S^{\dagger}(E)-V^{\dagger}(E)}
{\displaystyle \ss\cdot\p
        \Psi_{\textrm {f}+}} \end{array}\right) \ ,
\fineq
where $S(E)$ and $V(E)$ are the scalar and vector 
energy-dependent
components of the relativistic optical potential for a nucleon
with energy $E$ \cite{chc}. 
The upper component, $\Psi_{\textrm {f}+}$, is related to a
Schr\"odinger equivalent wave function, $\Phi_{\textrm{f}}$, by the Darwin 
factor, i.e.,
\inieq
\Psi_{\textrm {f}+} &=& \sqrt{D_{{\cal E}}^{\dagger}(E)}\Phi_{\textrm{f}} \ , \\
D_{{\cal E}}(E) &=& 1 + \frac{S(E)-V(E)}{M+{\cal E}} \ . \label{eq.darw}
\fineq
$\Phi_{\textrm{f}}$ is a two-component wave function which is solution of a 
Schr\"odinger
equation containing equivalent central and spin-orbit potentials obtained from
the scalar and vector potentials~\cite{clark,HPa}.

As no relativistic optical potentials are available for 
the bound states, then the wave function $\varphi_n$ is taken as 
the Dirac-Hartree solution of a relativistic Lagrangian
containing scalar and vector potentials \cite{adfx,lala}. 

Concerning the nuclear current operator, no unambiguous approach exists for 
dealing with off-shell nucleons. In the present calculations we use the cc2 
expression of the 1-body current \cite{defo,RDWIA1,meu}
\begin{eqnarray}
  j_{\textrm{cc}2}^{\mu} &=& F_1(Q^2) \gamma ^{\mu} + 
             i\frac {\kappa}{2M} F_2(Q^2)\sigma^{\mu\nu}q_{\nu} \ ,
	     \label{eq.cc}
\end{eqnarray}
where $\kappa$ is the anomalous part 
of the magnetic moment, $F_1$ and $F_2$ are the Dirac and Pauli nucleon form 
factors, which are taken from Ref.~\cite{mmd}, and
$\sigma^{\mu\nu}=\left(i/2\right)\left[\gamma^{\mu},\gamma^{\nu}\right]$.

Current conservation is restored by replacing the longitudinal current 
by \cite{defo}
\begin{eqnarray}
J^{\textrm {L}} &=& J^{\textrm {z}} = \frac{\omega}{\mid\q\mid}~J^0 \  .
\end{eqnarray}


\begin{figure}[ht]
\includegraphics[height=10cm, width=8.4cm]{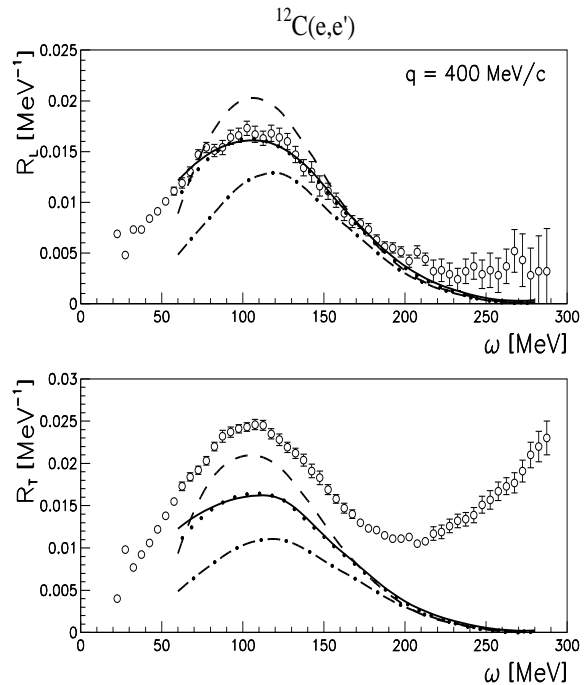} 
\vskip -0.5cm
\caption {Longitudinal (upper panel) and transverse (lower panel) response 
functions for the
$^{12}$C$(e,e')$ reaction at $q = 400$ MeV$/c$.  
Solid and dotted lines represent the NLSH results with and without the inclusion 
of the factor in Eq. (\ref{eq.def}), respectively. Dashed lines give the result 
without the integral in Eq. (\ref{eq.finale}).
Dot-dashed lines are the contribution of integrated single nucleon knockout
only. The data are from Ref. \cite{saclay}.}
\label{fig1}
\end{figure}


\begin{figure}[h]
\includegraphics[height=10cm, width=8.4cm]{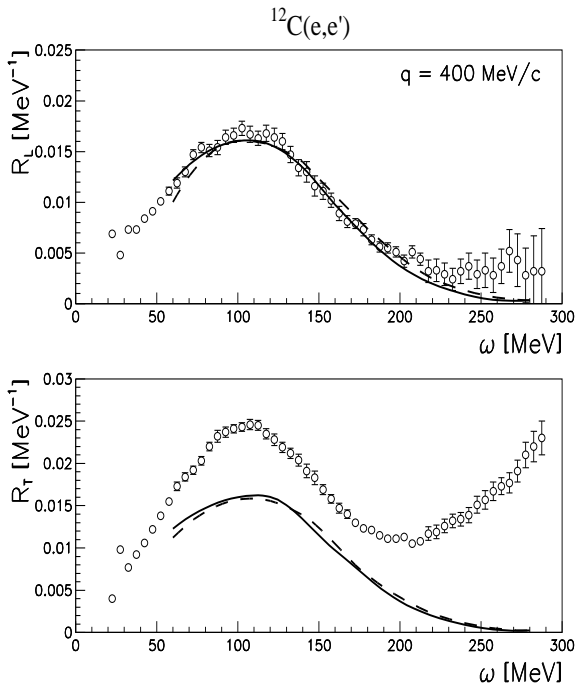} 
\vskip -0.5cm
\caption {Longitudinal (upper panel) and transverse (lower panel) response 
functions for the
$^{12}$C$(e,e')$ reaction at $q = 400$ MeV$/c$.   
Solid lines represent the NLSH results, dashed lines the NL3 results. Data as 
in Fig. \ref{fig1}.}
\label{fig1bis}
\end{figure}


\begin{figure}[h]
\includegraphics[height=10cm, width=8.4cm]{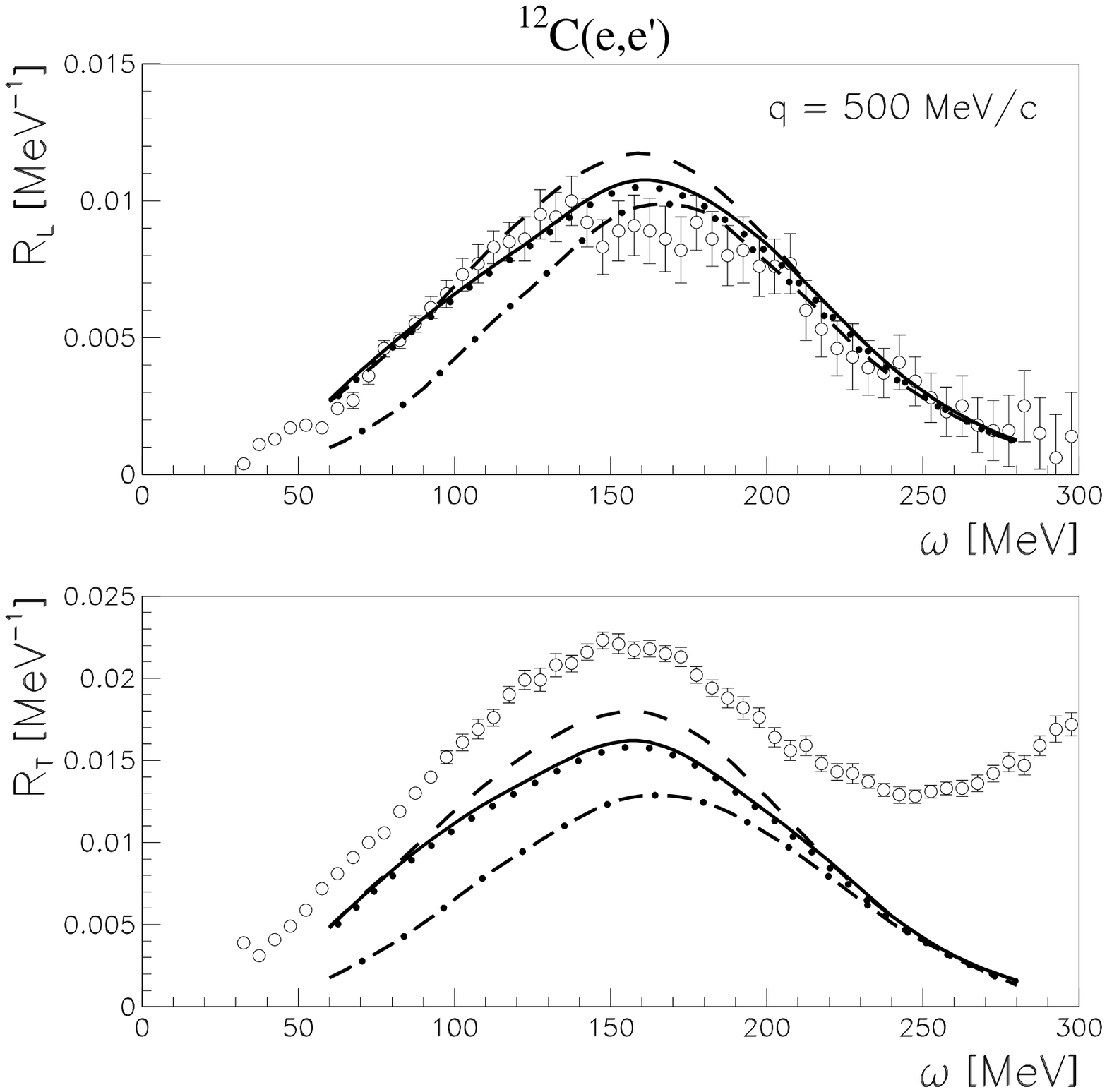} 
\vskip -0.5cm
\caption {The same as in Fig. \ref{fig1}, but for $q = 500$ MeV$/c$. The data
are from Ref. \cite{saclay}.}
\label{fig2}
\end{figure}


\begin{figure}[h]
\includegraphics[height=10cm, width=8.4cm]{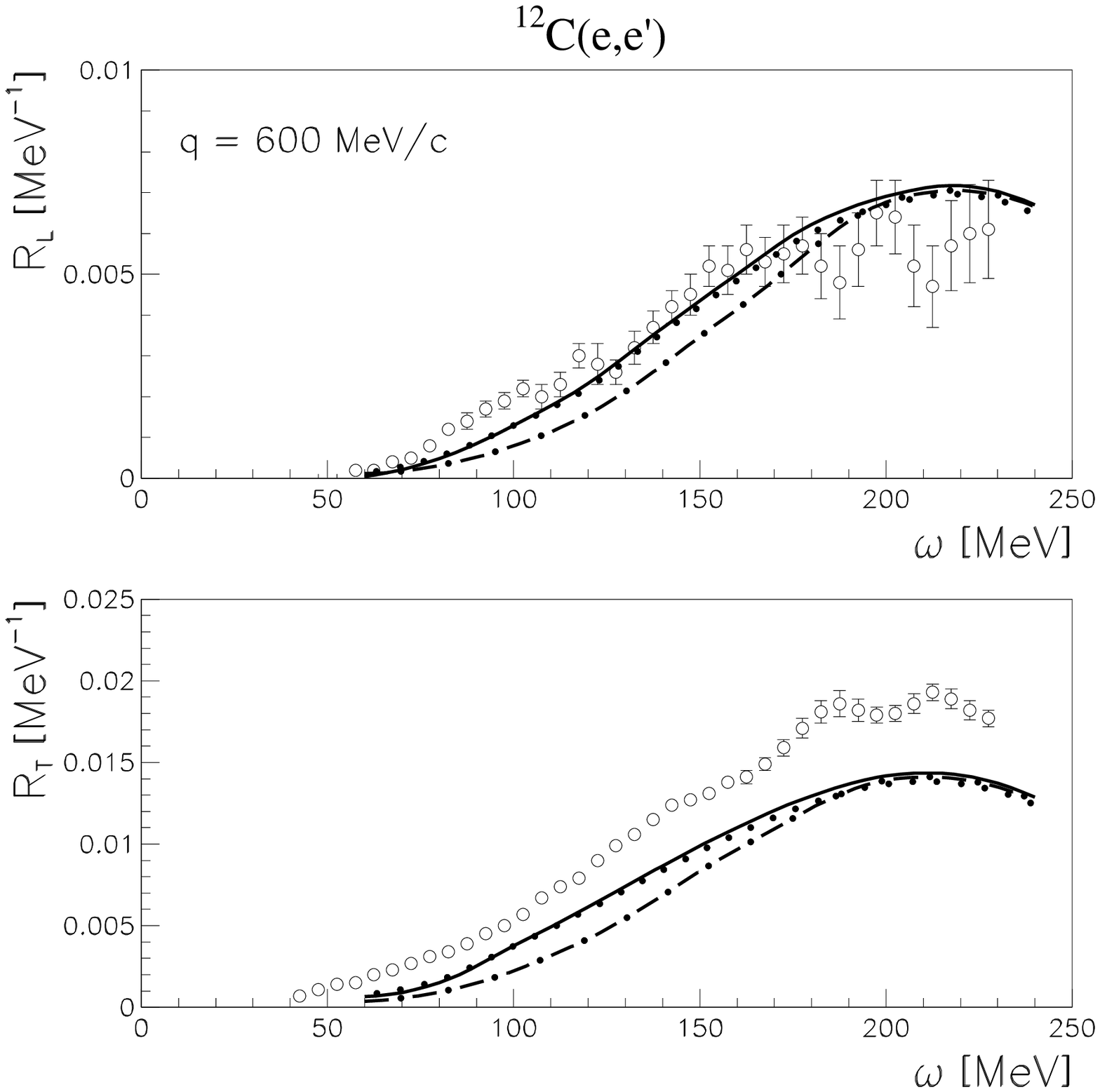} 
\vskip -0.5cm
\caption {The same as in Fig. \ref{fig1}, but for $q = 600$ MeV$/c$. The data
are from Ref. \cite{saclay}. }
\label{fig3}
\end{figure}

\begin{figure}[h]
\includegraphics[height=10cm, width=8.4cm]{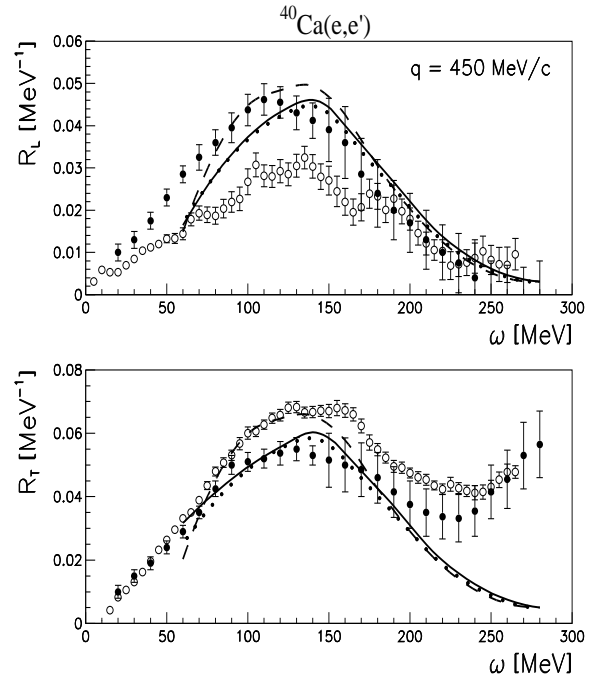} 
\vskip -0.5cm
\caption {Longitudinal (upper panel) and transverse (lower panel) response 
functions for the
$^{40}$Ca$(e,e')$ reaction at $q = 450$ MeV$/c$. The Saclay
data (open circles) are from Ref. \cite{meziani}, the MIT-Bates (black circles)
are from Ref. \cite{batesca}. Line convention as in Fig. \ref{fig1}.}
\label{fig4}
\end{figure}


\begin{figure}[h]
\includegraphics[height=10cm, width=8.4cm]{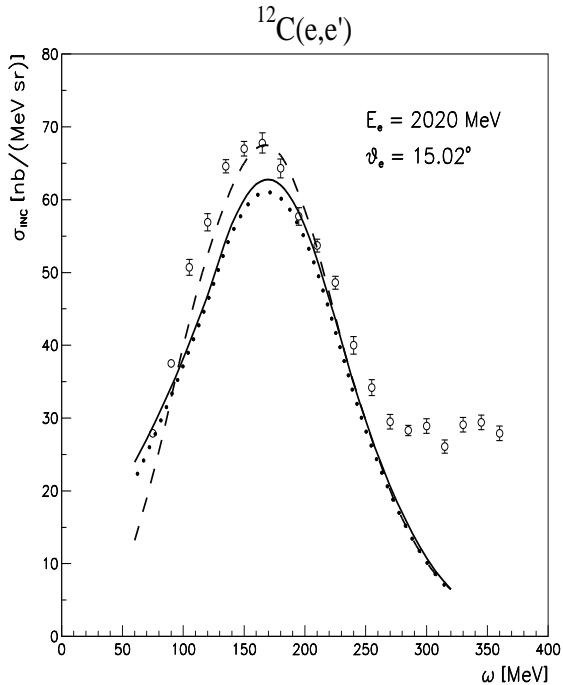} 
\vskip -0.5cm
\caption {The cross section for the inclusive $^{12}$C$(e,e')$ reaction at 
$\vartheta_e = 15.02^{\mathrm o}$ and $E_e = 2020$ 
MeV. The data are from SLAC \cite{csslac}. Line convention as in 
Fig. \ref{fig1}.}
\label{fig5}
\end{figure}


\begin{figure}[h]
\includegraphics[height=10cm, width=8.4cm]{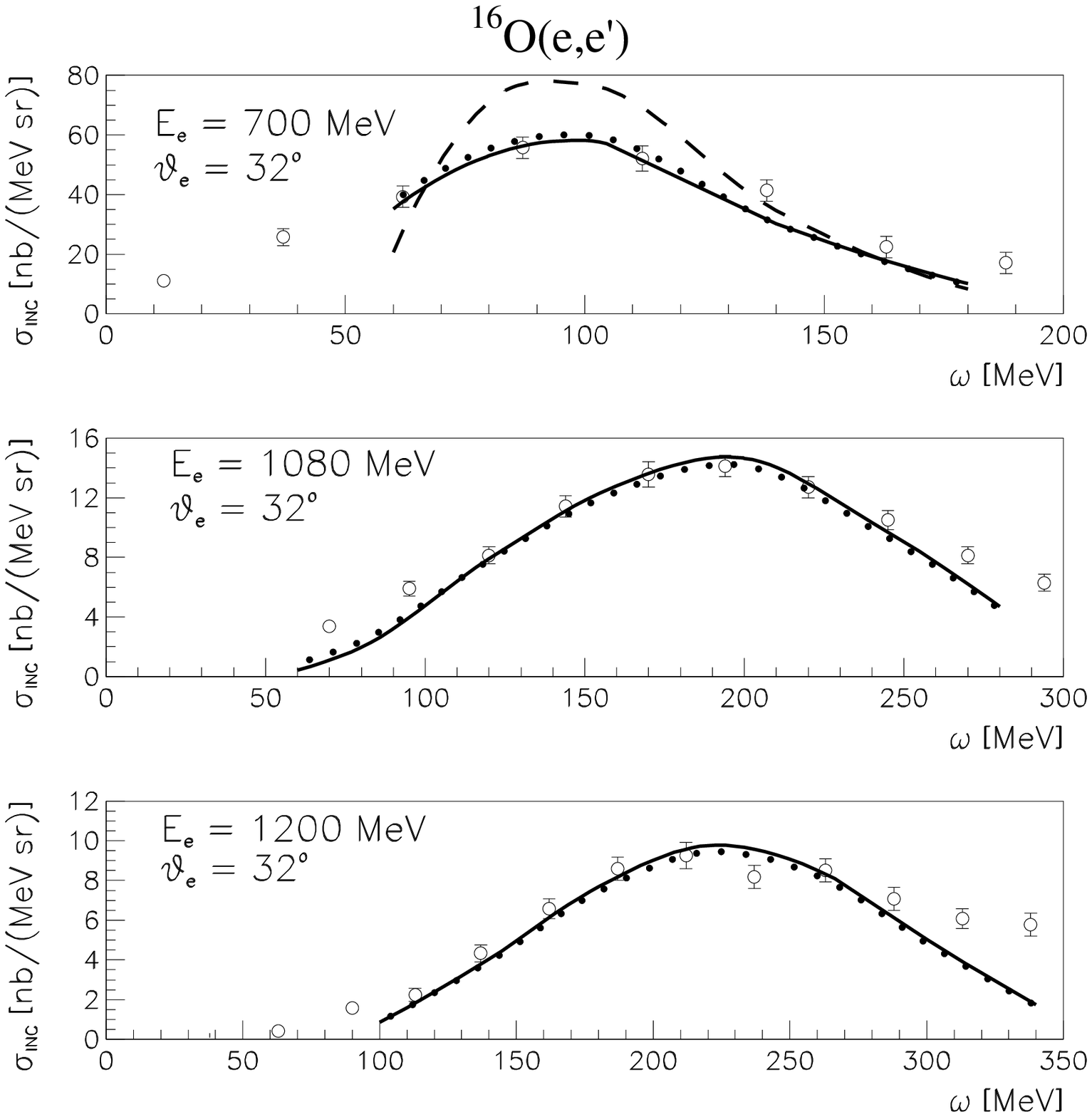} 
\vskip -0.5cm
\caption {The cross section for the inclusive $^{16}$O$(e,e')$ reaction at 
$\vartheta_e = 32^{\mathrm o}$ and $E_e = 700$, $1080$, and $1200$ MeV. 
The data are from ADONE-Frascati \cite{csfrascati}. Line convention as in 
Fig. \ref{fig1}.}
\label{fig6}
\end{figure}


The calculations have been performed with the same
bound state wave functions and optical potentials as in
Refs.~\cite{meucci1,meucci2}, where the RDWIA  was able to 
reproduce $\left(e,e^{\prime}p\right)$ and $\left(\gamma,p\right)$ 
data. 

The relativistic bound state wave functions have been obtained from the code of
Ref.~\cite{adfx}, where relativistic Hartree-Bogoliubov equations are solved in
the context of a relativistic mean field theory to reproduce
single-particle properties of several spherical and deformed nuclei~\cite{lala}. 
The scattering state is calculated by means of
the energy- and mass number-dependent EDAD1 complex phenomenological optical 
potential of Ref.~\cite{chc}, which is fitted to proton
elastic scattering data on several nuclei in an energy range up to 1040 MeV.

In the calculations the residual nucleus states $\mid n \rangle$ are restricted 
to be single particle one-hole states in the target. A pure shell model is 
assumed for the nuclear structure, i.e., we take a unitary spectral strength 
for each single particle state and the sum runs over all the occupied states.

The results presented in the following contain the contributions of
both terms in Eq. (\ref{eq.finale}). The calculation of the second term, which 
requires integration over all the eigenfunctions of the
continuum spectrum of the optical potential, is a rather complicate 
task. This term was
neglected in the nonrelativistic investigation of Ref.~\cite{capuzzi}, where its
contribution was estimated to be very small. In the present relativistic
calculations the effect of this term can be significant and it is therefore 
included in the results.

\subsection{The $R_{\textrm{L}}$ and $R_{\textrm{T}}$ response functions}

The longitudinal and transverse response functions for $^{12}$C at 
$q = 400$ MeV$/c$ are displayed in Fig. \ref{fig1} and compared with the Saclay 
data \cite{saclay}. The low energy 
transfer values are not given because the relativistic optical 
potentials are not available at low energies.

The agreement with the data is generally
satisfactory for the longitudinal response. 
In contrast, the transverse response is underestimated. This
is a systematic result of the calculations. It may be attributed to
physical effects which are not considered in the present approach, e.g., meson 
exchange currents. 

The effect of the integral in Eq. (\ref{eq.finale}) is also displayed.
At variance with the nonrelativistic result, here this contribution is 
important and essential to reproduce the experimental longitudinal response.

As explained in Sect. \ref{sec.single}, the contribution arising from 
interference between different channels, see Eqs. (\ref{eq.simple}) and 
(\ref{eq.approximate}), gives rise to the factor 
\inieq
 \sqrt{1- \mcv'_{\mathrm{L}}(E)}= \sqrt{1- \beta S'(E) - V'(E)} \ .
 \label{eq.def}
\fineq
 We 
see, however, that here it gives only a 
slight contribution, due to a compensation between the energy derivatives  
$S'(E)$ and $V'(E)$, while in the nonrelativistic calculation an overall 
reduction was observed, which was necessary to reproduce 
the data \cite{capuzzi}. 

The contribution from all the integrated single nucleon knockout
channels is also drawn in Fig. \ref{fig1}. It is significantly smaller than the 
complete calculation. The reduction, which is larger at lower values of
$\omega$, gives an indication of the relevance of inelastic channels.  

For the calculations in Fig. \ref{fig1} the Hartree-Bogoliubov equations for 
the single particle bound states have been solved using NLSH choices for the 
parameters of the relativistic mean field theory Lagrangian. 
In Fig. \ref{fig1bis} a comparison is shown between the results obtained with
this choice of parameters and a different choice, i.e., NL3.
The shapes of the responses calculated with the different bound states show
small differences. Their integrals must be unchanged, according to the fact
that the sum rule has to be preserved. 

The longitudinal and transverse response functions for $^{12}$C at $q = 500$ 
and $q = 600$ MeV$/c$ are displayed in Figs. \ref{fig2} and \ref{fig3},
respectively, and compared with the Saclay data \cite{saclay}. 
The bound state wave function have been obtained with the NLSH parametrization. 
As already found in Fig. \ref{fig1} at $q = 400$ MeV$/c$, a good agreement with 
the data is obtained in both cases for the longitudinal response, while the 
transverse response is underestimated. 
Also in Figs. \ref{fig2} and \ref{fig3} only a slight effect is given by the 
factor in Eq. (\ref{eq.def}) arising from the interference between different
channels. The role of the integral in Eq. (\ref{eq.finale}) decreases 
increasing 
the momentum transfer. At $q = 500$ MeV$/c$ its contribution is smaller than 
at $q = 400$ MeV$/c$, but still important to reproduce the experimental 
longitudinal response, while at  $q = 600$ MeV$/c$ the effect is negligible and
the two curves with and without the integral overlap. The effect of the 
inelastic channels, indicated in the figures by the difference between the 
complete results and the contribution from all the integrated single nucleon 
knockout channels, is always visible and even sizable, but it decreases 
increasing the momentum transfer.

The response functions for $^{40}$Ca at $q = 450$ MeV$/c$ 
are shown in Fig. \ref{fig4} and compared with the Saclay \cite{meziani} and
the MIT-Bates \cite{batesca} data. 
The results obtained with the NLSH set of parameters have been plotted, 
since the results with other sets are almost equivalent. 
The calculated response functions are of the same order of magnitude as the 
MIT-Bates data, while for the Saclay data the longitudinal response is 
overestimated and the transverse response underestimated. 
The factor in Eq. (\ref{eq.def}) produces and enhancement which is minimal but
visible in the figure.

\subsection{The inclusive cross section}
 
Investigation of inclusive electron scattering in the region of large $q$ is of
great interest to provide information on the nuclear wave functions and
excitation and decay of nucleon resonances. Several experiments have been
carried out to explore this region. The separation of the longitudinal and
transverse components of the nuclear response would be very interesting, but it 
is very difficult to perform because of the decreasing of the 
longitudinal-transverse ratio with increasing $q$. Precise measurements over a 
kinematical range that would allow longitudinal-transverse separation for 
several nuclei are however planned in the future at JLab, where the E-01-016 
approved experiment \cite{jlabpro} will make a precise measurement in the
momentum transfer range $0.55 \le q \le 1.0$ GeV$/c$ in order to extract the
response functions.

In this Subsection we focus our attention on experimental cross sections 
with $\omega \lesssim 300$ MeV, since our model does not include meson 
exchange currents and isobar excitation contributions. 

The calculated inclusive $^{12}$C$(e,e')$ cross section is displayed in 
Fig. \ref{fig5} in comparison with the SLAC data \cite{csslac} in a kinematics 
with a beam energy $E_e = 2020$ MeV and a scattering angle of 
$\simeq 15^{\mathrm o}$. The bound state wave function has been obtained with 
the NLSH set. A visible enhancement is produced by the factor in 
Eq. (\ref{eq.def}). The effect of the integral in Eq. (\ref{eq.finale}) 
gives a significant reduction which underestimates the data. As in the case of
the transverse response of Figs. \ref{fig1} - \ref{fig4}, the discrepancy 
might be due to two-body mechanisms which are here neglected.

In order to extend our analysis to different kinematics and target nuclei, we 
consider in Fig. \ref{fig6} the $^{16}$O$(e,e')$ inclusive cross section data 
taken at ADONE-Frascati \cite{csfrascati} with beam energy ranging from $700$ 
to $1200$ MeV and a scattering angle of $\simeq 32^{\mathrm o}$. The NLSH wave 
functions have been used in the calculations. The agreement with data is 
good in all the considered situations. The integral in Eq. (\ref{eq.finale}) 
produces a 
reduction which is now essential to reproduce the data at $700$ MeV, which 
correspond to a momentum transfer of $\lesssim 400$ MeV$/c$. Its
contribution can be neglected in the other kinematics, where $q \simeq 600$ 
MeV$/c$. The effect of the factor in Eq. (\ref{eq.def}) is very small.

\section{Summary and conclusions}
\label{conc}

A relativistic approach to inclusive electron scattering in the
quasielastic region has been presented. This work can be considered as an 
extension of the nonrelativistic many-body approach of Ref. \cite{capuzzi}. 
The components of the hadron tensor are written in terms of Green functions of
the optical potentials related to the various reaction channels. The projection
operator formalism is used to derive this result. An explicit calculation of
the single particle Green function can be avoided by means of its spectral 
representation, based on a biorthogonal expansion in terms of the 
eigenfunctions of 
the nonhermitian optical potential $\mcv(E)$ and of its hermitian conjugate. The
interference between different channels is taken into account by the factor
$\sqrt{1-\mcv'(E)}$, which also allows the replacement of the mean 
field $\mcv(E)$ by the phenomenological optical potential 
$\mcv_{\textrm{L}}(E)$. After this replacement, the nuclear response functions 
are expressed in terms 
of matrix elements similar to the ones which appear in the exclusive one
nucleon knockout reactions, and the same RDWIA treatment \cite{meucci1} can be
applied to the calculation of the inclusive electron scattering. 

The effects of final state interactions are thus described consistently in 
exclusive and inclusive processes. Both the real and imaginary parts of the 
optical potential must be included. In the exclusive reaction the imaginary 
part accounts for the flux lost towards other final states. In the inclusive
reaction, where all the final states are included, the imaginary part accounts
for the redistribution of the strength among the different channels. 

All the final states contributing to the inclusive reaction are contained in 
the Green function, and not only one nucleon emission. Our calculations for the 
inclusive electron scattering are different from the contribution of integrated 
single nucleon knockout only. The difference between the two results is
originated by the imaginary part of the optical potential.

The transition matrix elements are calculated using the bound state wave 
functions obtained in the framework of a relativistic mean field theory. The 
direct Pauli reduction method is applied to the scattering wave functions.

Numerical results for the longitudinal and transverse response functions of 
$^{12}$C and $^{40}$Ca have been presented in comparison with data in a
momentum transfer range between 400 and 600 MeV$/c$.  

The role and relevance of the various effects of final state interactions can 
be different in the relativistic and nonrelativistic calculations. This is a
consequence of the different features of the optical potentials in the two
approaches. The final effect is however similar and produces qualitatively
similar results in comparison with data. The relativistic frame has anyhow 
the advantage that it can more reliably be applied to a wider range 
of situations and kinematics.

Our relativistic results confirm that the effects of final state interactions 
are large and essential to reproduce the data. 
The term with the integral, entering the definition of the hadron tensor 
$W^{\mu\mu}(\omega,q )$ in Eq. (\ref{eq.finale}), gives a significant 
contribution, which is important to improve the agreement with data. This
result is different from the one obtained in the nonrelativistic analysis 
\cite{capuzzi}, where this term gave only a small contribution and was thus 
neglected in the calculations. We stress that this term is due to the 
imaginary part of the optical potential, which thus produces different but 
important effects in the relativistic and nonrelativistic approaches. 

The effects of the integral in Eq. (\ref{eq.finale}) as well as the difference 
between the 
complete result and the contribution of integrated single nucleon knockout, 
which are both entirely due to the imaginary part of the optical potential, 
tend to decrease with increasing momentum transfer. 

The factor $\sqrt{1-\mcv'(E)}$ is conceptually very important. 
It accounts for interference effects and allows the replacement of 
$\mcv(E)$ by $\mcv_{\textrm{L}}(E)$. In the nonrelativistic analysis of 
Ref.~\cite{capuzzi} this factor produced an overall reduction of 
the calculated strength which significantly improved the agreement with the 
experimental longitudinal response function. Only a small contribution is 
given by this factor in the present relativistic approach. It generally 
produces a small enhancement of the calculated responses that does not change
significantly the comparison with data. 

Final state interactions have a similar effect on the longitudinal and
transverse components of the nuclear response. In comparison with data, the 
longitudinal response is usually well reproduced, while the transverse response 
is underestimated. This seems to indicate that more complicated effects, e.g., 
two-body meson exchange currents, have to be added to the present single 
particle approach.

The inclusive cross section for $^{12}$C and $^{16}$O has been calculated for
momentum transfer $\lesssim$ 600 MeV$/c$. The results for $^{12}$C are in 
agreement with  those obtained for the response functions. The lack of strength 
in the determination of the transverse response results in an underestimation 
of the data. A satisfactory agreement is obtained for the $^{16}$O$(e,e')$ 
results.




\begin{thebibliography}{}
\bibitem{book}
S. Boffi, C. Giusti, F. D. Pacati, and M. Radici,
{\it Electromagnetic Response of Atomic Nuclei}, Oxford Studies in Nuclear
Physics, Vol. 20 (Clarendon Press, Oxford, 1996).

\bibitem{csfrascati}
M. Anghinolfi {\it et al.},
Nucl. Phys. {\bf A602}, 405 (1996).

\bibitem{batesca}
C.F. Williamson {\it et al.},
Phys. Rev. C {\bf 56}, 3152 (1997).


\bibitem{jlabpro}
J.P. Chen, S. Choi, and Z.E. Meziani, spokespersons, JLab experiment E-01-016.

\bibitem{Sluys}
V. Van der Sluys, J. Ryckebusch, and M. Waroquier,
Phys. Rev. C {\bf 51}, 2664 (1995).

\bibitem{Cenni}
R. Cenni, F. Conte, and P. Saracco,
Nucl. Phys. {\bf A623}, 391 (1997).

\bibitem{Amaro}
J.E. Amaro, M.B. Barbaro, J.A. Caballero, T.W. Donnelly, and A. Molinari,
Phys. Rep. {\bf 368}, 317 (2002); nucl-th/0301023.

\bibitem{Fabrocini}
A. Fabrocini,
Phys. Rev. C {\bf 55}, 338 (1997)

\bibitem{Co}
G. Co' and A.M. Lallena, Ann. Phys. {\bf 287}, 101 (2001).

\bibitem{Leidemann}
W. Leidemann and G. Orlandini,
Nucl. Phys. {\bf A506}, 447 (1990)

\bibitem{Sick}
I. Sick,
in {\it Nuclear Theory}, (Heron Press Science Series, Sofia, 2002), p. 16.

\bibitem{hori}
Y. Horikawa, F. Lenz, and N.C. Mukhopadhyay,
Phys. Rev. C {\bf 22}, 1680 (1980).

\bibitem{chinn}
C.R. Chinn, A. Picklesimer, and J.W. Van Orden,
Phys. Rev. C {\bf 40}, 790 (1989); Phys. Rev. C {\bf 40}, 1159 (1989).

\bibitem{bouch}
P.M. Boucher and J.W. Van Orden,
Phys. Rev. C {\bf 43}, 582 (1991).

\bibitem{capuzzi}
F. Capuzzi, C. Giusti, and F.D. Pacati,
Nucl. Phys. {\bf A524}, 681 (1991).

\bibitem{capma}
F. Capuzzi and C. Mahaux,
Ann. Phys. (N.Y.) {\bf 254}, 130 (1997).

\bibitem{west}
G.B. West,
Phys. Rep. {\bf 18}, 264 (1975).


\bibitem{orlandini}
G. Orlandini and M. Traini,
Rept. Prog. Phys. {\bf 54}, 257 (1991).


\bibitem{fesh}
H. Feshbach, 
Ann. Phys. (N.Y.) {\bf 5}, 357 (1958).

\bibitem{bofficapuzzi}
S. Boffi and F. Capuzzi,
Nucl. Phys. {\bf A351}, 219 (1981).

\bibitem{capuzzi2} 
F. Capuzzi,
Nucl. Phys. {\bf A554}, 362 (1993).

\bibitem{fesh2}
H. Feshbach, 
Ann. Phys. (N.Y.) {\bf 19}, 287 (1962).

\bibitem{capma2}
F. Capuzzi and C. Mahaux,
Ann. Phys. (N.Y.) {\bf 239}, 57 (1995).

\bibitem{capma3}
F. Capuzzi and C. Mahaux,
Ann. Phys. (N.Y.) {\bf 245}, 147 (1996).

\bibitem{meucci1}
A. Meucci, C. Giusti, and F. D. Pacati, 
 Phys. Rev. C {\bf  64}, 014604 (2001).

\bibitem{meucci2}
A. Meucci, C. Giusti, and F. D. Pacati, 
 Phys. Rev. C  {\bf  64}, 064615 (2001).

\bibitem{chc} E.D. Cooper, S. Hama, B.C. Clark, and R.L. Mercer,
 Phys. Rev. C {\bf 47}, 297 (1993).

\bibitem{clark} B.C. Clark,
    in {\sl Proceedings of the Workshop on Relativistic Dynamics and
   Quark-Nuclear Physics}, edited by M.B. Johnson and A. Picklesimer 
   (John Wiley \& Sons, New York, 1986), p. 302. 

\bibitem{HPa}
M. Hedayati-Poor, J.I. Johansson, and H.S. Sherif,
 Nucl. Phys. {\bf A593}, 377 (1995).

\bibitem{adfx}
W. P\"oschl, D. Vretenar, and P. Ring, 
Comput. Phys. Commun. {\bf 103}, 217 (1997).

\bibitem{lala}
G.A. Lalazissis, J. K\"onig, and P. Ring, 
Phys. Rev. C {\bf 55}, 540 (1997).

\bibitem{RDWIA1}
J.J. Kelly,
 Phys. Rev. C {\bf  60}, 044609 (1999).

\bibitem{defo}
T. de Forest, Jr., 
Nucl. Phys. {\bf A392}, 232 (1983).

\bibitem{meu}
A. Meucci,  
 Phys. Rev. C  {\bf  65}, 044601 (2002).

\bibitem{mmd} 
P. Mergell, Ulf-G. Meissner, and D. Drechsel,
 Nucl. Phys. {\bf A596}, 367 (1996).

\bibitem{saclay}
P. Barreau {\it et al.},
Nucl. Phys. {\bf A402}, 515 (1983); Note CEA N-2334.

\bibitem{meziani}
Z.E. Meziani {\it et al.},
Phys. Rev. Lett. {\bf 52}, 2130 (1984); {\bf 54}, 1233 (1985).

\bibitem{csslac}
D.B. Day {\it et al.},
Phys. Rev. C { \bf 48}, 1849 (1993).

\end{thebibliography}
\end{document}